\documentstyle[12pt]{article}
\begin{document}

\tolerance=5000

\def\pp{{\, \mid \hskip -1.5mm =}}
\def\cL{{\cal L}}
\def\be{\begin{equation}}
\def\ee{\end{equation}}
\def\bea{\begin{eqnarray}}
\def\eea{\end{eqnarray}}
\def\tr{{\rm tr}\, }
\def\nn{\nonumber \\}
\def\e{{\rm e}}
\def\D{{D \hskip -3mm /\,}}

\def\SEH{S_{\rm EH}}
\def\SGH{S_{\rm GH}}
\def\AdS5{{{\rm AdS}_5}}
\def\S4{{{\rm S}_4}}
\def\gfv{{g_{(5)}}}
\def\gfr{{g_{(4)}}}
\def\SC{{S_{\rm C}}}
\def\RH{{R_{\rm H}}}

\def\wlBox{\mbox{
\raisebox{0.1cm}{$\widetilde{\mbox{\raisebox{-0.1cm}\fbox{\ }}}$}}}
\def\htBox{\mbox{
\raisebox{0.1cm}{$\hat{\mbox{\raisebox{-0.1cm}{$\Box$}}}$}}}

\  \hfill
\begin{minipage}{3.5cm}
May 2002 \\
\end{minipage}

\vfill

\begin{center}
{\large\bf Entropy bounds and Cardy-Verlinde formula in Yang-Mills theory}

\vfill

{\sc Shin'ichi NOJIRI}\footnote{nojiri@cc.nda.ac.jp},
and {\sc Sergei D. ODINTSOV}$^{\spadesuit}$\footnote{
odintsov@mail.tomsknet.ru}\\

\vfill

{\sl Department of Applied Physics \\
National Defence Academy,
Hashirimizu Yokosuka 239-8686, JAPAN}

\vfill

{\sl $\spadesuit$ 
Lab. for Fundamental Study,
Tomsk State Pedagogical University \\
634041 Tomsk, RUSSIA}

\vfill

{\bf ABSTRACT}

\end{center}
Using gauge formulation of gravity the three-dimensional $SU(2)$ 
YM theory equations of motion are presented in equivalent form 
as FRW cosmological equations. With the radiation, the particular 
(periodic, big bang-big crunch) three-dimensional universe is constructed.
Cosmological 
entropy bounds (so-called Cardy-Verlinde formula) have the standard form
in such universe. Mapping such universe back to YM formulation 
we got the thermal solution of YM theory. The corresponding holographic
entropy bounds (Cardy-Verlinde formula) in YM theory are constructed.
This indicates to universal character of holographic relations.

\newpage

It becomes popular in theoretical physics to apply 
different formulations (and even theories) in the 
description of the same phenomenon. For example,
it is expected that Einstein theory presented in Yang-Mills (YM) 
form is easier to quantize. Moreover, its renormalizability 
properties seem to be better in YM-like form \cite{mitya}.
 From another point, there exists the YM theory presentation 
in Einstein-like form. In this relation the natural question is:
can one achieve some new results in YM theory using 
recent studies of cosmology based on holographic principle?

In this paper we start the investigation in this direction.
Using three-dimensional YM theory as an example we first 
rewrite it in the gravitational form. The YM equations of motion
are then rewritten as FRW cosmological equations where 
natural definition of Hubble, Bekenstein and Bekenstein-Hawking 
entropies  
maybe done. As a result FRW equation is represented in
the form similar to two-dimensional CFT entropy (so-called 
Cardy-Verlinde formula \cite{EV}). Introducing the radiation
(matter) the explicit solution of such three-dimensional periodic universe
is obtained. The cosmological entropy bounds (holographic Cardy-Verlinde
formula)
are quite simple for such universe. In particular, the Bekenstein entropy 
is constant. Such three-dimensional universe is then mapped back to YM 
theory form.
(Such procedure maybe considered as an interesting way
of generating of YM solutions from FRW cosmological solutions).
 The correspondent thermal YM solution on which YM action diverges 
is constructed. The divergence of action is absorbed into the renormalization 
of the gauge coupling constant. After the identification of YM entropy 
with Bekenstein entropy the Cardy-Verlinde formula defines the entropy
bounds in YM theory.

The action of Yang-Mills theory with the gauge group $SU(2)$ 
is given by
\bea
\label{Y1}
S_{\rm YM}&=&{1 \over 4g_{\rm YM}^2}\int d^dx \sum_{a=1}^3\left(
F^a_{ij}F^{a\,ij} + 4g_{\rm YM}^2A^a_i J^{ai}\right) \\
F^a_{ij}&=&\partial_i A^a_j - \partial_i A^a_j 
-\sum_{b,c=1}^3 \epsilon^{abc}A^b_i A^c_j\ .
\eea
Here $d$ is the dimension of the spacetime and 
$J^{ai}$ is the external source.  For
$d=3$,  one can rewrite the action (\ref{Y1}) in the following 
form:
\be
\label{Y2}
S_{\rm YM}={1 \over 2g_{\rm YM}^2}\int d^dx \sum_{a=1}^3\left(
B^a_iB^{a\,i} + 2g_{\rm YM}^2 A^a_i J^{ai}\right) \ .
\ee
Here $B^a_i$ is defined by 
\be
\label{II}
B^{ak}=\epsilon^{kij}\left(\partial_i A^a_j - {1 \over 2} 
\epsilon^{abc} A^b_i A^c_j\right)\ ,
\ee
The equations of motion are 
\be
\label{Y3}
0=\epsilon^{kij}D_k B^a_j - g_{\rm YM}^2 J^{ai}
=\epsilon^{kij}\left(\partial_k B^a_j - {1 \over 2} 
\epsilon^{abc} A^b_k B^c_j\right) - g_{\rm YM}^2 J^{ai}\ .
\ee
Here $D_i$ expresses the covariant derivative. On the 
other hand,  Bianchi identities are:
\be
\label{Y4}
0=D_i B^{a\,i}\ .
\ee

There  exists a formulation of gravity starting from the gauge theory, 
whose gauge group is the Poincare group composed of translations 
and Lorentz transformations. The gauge field of the translation 
corresponds to the vierbein (or vielbein)  and that of the 
Lorentz tansformation to the spin-connection. The gauge curvature 
of the translation corresponds to the torsion. Then 
if we impose a condition that the curvature vanishes, we can solve 
the spin connection in terms of the vier(viel)bein field if the 
dimension of the spacetime is larger than 2. As the Lorentz group 
is $S(d-1,1)$ for $d$-dimensional spacetime,  one can rewrite the 
gauge field of $SO(d-1,1)$ by the vier(viel)bein field. 
Such a formulation for the supergravity is developed in \cite{KU}. 
 In the similar way in ref.\cite{H},
 4 dimensional Yang-Mills theory is presented 
in the Hamiltonian formulation in terms of 3 dimensional 
gravity\footnote{The presentation of SUSY YM theory in
terms of Riemann geometry is recently considered 
in \cite{Ricardo}.}. 
The parametrization of the gauge field used in the paper 
corresponds to the expression of the spin connection by the 
dreibein fields. 

In the following, as a most simple and non-trivial case, we consider 
the case that the dimension of the spacetime is 3 ($d=3$) and 
the spacetime is Euclidean, where the Lorentz group is $SO(3)$ or 
$SU(2)$. 
In case that the spacetime is 3 dimensional Euclidean space, 
the Poincare algebra is generated by the generators of 
the translation $T^a$ $(a=1,2,3)$ and the rotation $R^a$ $(a=1,2,3)$ 
around $a$-axis. The algebra is explicitly 
given by 
\be
\label{alg}
[R^a, R^b]=\epsilon^{abc}R^c\ ,\quad 
[R^a, T^b]=\epsilon^{abc}T^c\ ,\quad [T^a, T^b]=0\ .
\ee
Denote the gauge fields corresponding to the 
translation and the rotation by $u^a_i$ and $A^a_i$ 
($a=1,2,3$), respectively. Then the curvatures $R_{ij}^{Ta}$ 
corresponding to the translation and $R_{ij}^{Ra}$ 
corresponding to the rotation are given by
\bea
\label{curvG}
R_{ij}^{Ta}&=&\partial_i u_j^a - \partial_i u_j^a 
 - \epsilon^{abc} \left(A_i^b u_j^c - A_j^b u_i^c\right) \ ,\nn
R_{ij}^{Ra}&=&\partial_i A_j^a - \partial_i A_j^a 
 - \epsilon^{abc} A_i^b A_j^c \ .
\eea
The  relation
\be
\label{tor}
R_{ij}^{Ta}=0
\ee
corresponds to the torsionless condition in gravity. 
The condition (\ref{tor}) can be solved with respect to 
$A_i^a$: 
\bea
\label{I}
A_i^a&=&-\epsilon^{abc}\left(\partial_i u^b_j 
 - \partial_j u^b_i\right)\left(u^{-1}\right)_c^j 
+ {1 \over 4}\epsilon^{bcd}\left(\partial_k u^b_j 
 - \partial_j u^b_k\right)\left(u^{-1}\right)_c^j
 \left(u^{-1}\right)_d^k u^a_i \nn
&=&{\epsilon^{nmk}\partial_m u^b_k\left(u^a_n u^b_i - 
{1 \over 2}u^b_n u^a_i\right) \over \det u}\ .
\eea
Then the field strength $B^{ak}$ defined by (\ref{II}) 
is related with the Einstein tensor $G^{ij}=R^{ij} - 
{1 \over 2}g^{ij}R$ by
\be
\label{III}
B^{ai}=\sqrt{g}u^a_j G^{ij}\ .
\ee
Here the metric tensor $g_{ij}$ is defined by 
\be
\label{IV}
g_{ij}=u^a_i u^b_j 
\ee
and $R^{ij}$ and $R$ are the Ricci curvature and the 
scalar curvature constructed on $g_{ij}$. Let denote the 
covariant derivative with respect to the $SO(3)$ gauge 
group by $D_i$ and the covariant derivative with 
respect to the gravity by $\nabla_i$. Then Eq.(\ref{III}) 
tells that 
\be
\label{IVb}
D_j B^{ai}=\sqrt{g}u^a_j \left(\nabla_j G^{ij}
 - \Gamma^i_{jl}G^{kl} + \Gamma^l_{lj} G^{kl}\right)\ ,
\ee
which leads
\be
\label{V}
D_i B^{ai}=\sqrt{g}u^a_j \nabla_i G^{ij}\ .
\ee
The Bianchi identity
\be
\label{VI}
 \nabla_i G^{ij}=0 
\ee
can be consistent with the gauge Bianchi identity 
(\ref{Y4}). 
By using (\ref{III}), the 
Einstein equation 
\be
\label{VII}
G^{ij}=\kappa^2 T^{ij}\ ,
\ee
with the energy-momentum tensor $T_{\mu\nu}$ 
can be rewritten in terms of the gauge fields:
\be
\label{VIII}
B^a_i=B^{{\rm ext} a}_i\ ,\quad 
B^{{\rm ext} a}_i\equiv {\kappa^2 \over2}
\sqrt{g}u^a_j T^{ij}\ .
\ee
Here  one can regard $B^{{\rm ext} a}_i$ as an external filed, 
which is related with $J^{ai}$ (\ref{Y3}) by
\be
\label{VIIIb}
\epsilon^{kij}D_k B^{{\rm ext} a}_j = g_{\rm YM}^2 J^{ai}\ . 
\ee
Since the conservation of the energy-momentum tensor 
$\nabla_i T^{ij}=0$ corresponds to the ``conservation law'' 
of the external fields, if the energy-momentum conservation 
holds one can construct the solution of the Bianchi identity 
(\ref{Y4}) from the solution of the Einstein 
equation (\ref{VII})  using (\ref{III}). 

Now we consider more concrete solution.  Assume the 
dreibein fields have a form of 
\be
\label{IX}
u^1_t=1\ ,\quad u^2_\theta=\e^{A(t)}\ ,\quad 
u^3_\phi =\sin\theta \e^{A(t)}\ ,
\quad \mbox{other components}=0\ ,
\ee
which give the Euclidean FRW metric
\be
\label{X}
ds^2=dt^2 + l^2\e^{2A(t)}\left(d\theta^2 + \sin^2\theta d\phi^2
\right)\ .
\ee
The Minkowski signature metric can be obtained  changing
the compact gauge group $SU(2)$ or $SO(3)$ to the non-compact 
ones $SU(1,1)$, $SO(2,1)$ or $SL(2,C)$. 
The assumptions (\ref{IX}) give
\bea
\label{XI}
&& A^1_t=A^2_t=A^3_t=A^1_\theta=A^2_\theta=A^3_\phi=0 \nn
&& A^3_\theta = l\e^A \dot A\ ,\quad 
A^2_\phi = - l\sin\theta \e^A \dot A\ , \quad
A^1_\phi = \cos\theta \ ,\\
\label{XII}
&& B^{2t}=B^{3t}=B^{1\theta}=B^{3\theta}=B^{1\phi}=B^{2\phi}=0 \nn 
&& B^{1t}=-l^2 \sin\theta \e^{2A}{\dot A}^2 - \sin\theta\ ,\quad 
B^{2\theta}=l\sin\theta\left(\ddot A + {\dot A}^2\right)\e^A\ ,\nn
&& B^{3\phi}=l\left(\ddot A + {\dot A}^2\right)\e^A\ .
\eea
If we define $B^{ab}$ by 
\be
\label{Bab}
B^{ab}=u^b_i B^{ai}\ ,
\ee
Eq.(\ref{III}) tells $B^{ab}$ is symmetric:
\be
\label{Bab2}
B^{ab}=B^{ba}\ .
\ee
 Any $N\times N$ symmetric matrix $M$, in general, 
can be diagonalized by a matrix $O$  from $SO(N)$, 
$M\to O^{-1}MO=\left(\begin{array}{ccccc} m_1 & 0 & & \cdots & 0 \\
0 & m_2 & 0 & \cdots & 0 \\
0 & 0 & \ddots & & \vdots \\
\vdots & \vdots & & \ddots & 0 \\
0 & 0 & \cdots & 0 & m_N \\
\end{array} \right)$. 
Then  using $SO(3)$ gauge transformation,  one can diagonalize 
$B^{ab}$.  The gauge transformation removes 3 components in 
$B^{ab}$, which corresponds to the degree of the freedom. 
 Eq.(\ref{IX}) and also 
Eq.(\ref{XII})  manifest such a special gauge choice.  

The stress-energy tensor may be presented in terms of the energy 
density $\rho$ and the pressure $p$ as
\be
\label{XIII}
T^{tt}=-\rho\ ,\quad T^{\theta\theta}=l^{-2}\e^{-2A}p\ ,\quad 
T^{\phi\phi}={1 \over l^2\sin\theta}\e^{-2A}p\ ,\quad 
\mbox{other components}=0\ ,
\ee
Thus
\bea
\label{XIV}
&& B^{{\rm ext}\,2t}=B^{{\rm ext}3t}=B^{{\rm ext}\,1\theta}
=B^{{\rm ext}\,3\theta}=B^{{\rm ext}\,1\phi}
=B^{{\rm ext}\,2\phi}=0 \\ 
&& B^{{\rm ext}\,1t}=-{\kappa^2 \over 2}
l^2\sin\theta\e^{2A} \rho\ ,\quad 
B^{{\rm ext}\,2\theta}={\kappa^2 \over 2}l\sin\theta\e^A p\ ,
\quad B^{{\rm ext}\,3\phi}={\kappa^2 \over 2}l\e^Ap\ .\nonumber
\eea
Then by combining (\ref{VIII}), (\ref{XI}) and (\ref{XIV}), 
we obtain the FRW equations:
\bea
\label{XV}
&& \dot A^2 + {1 \over a^2}={\kappa^2 \over 2}\rho \nn
&& \ddot A + \dot A^2={\kappa^2 \over 2}p \ .
\eea
If we define $J^{ai}$ by (\ref{VIIIb}), any solution of 
(\ref{XV}) satisfies the equations of motion (\ref{Y3}) 
by the identification  (\ref{XII}). 
 Moreover, if the energy momentum tensor (\ref{XIII}) satisfies 
the conservation law $\nabla_i T^{ij}=0$, the solution of 
the equation (\ref{XV}) also satisfies the Bianchi identity 
(\ref{Y4}). 

Here $a=l\e^A$. Combining (\ref{XI}) and (\ref{XIV}) with 
(\ref{VIIIb}), we may obtain the external source $J^{ai}$.

 In gravity one can define 
the Hubble, Bekenstein and Bekenstein-Hawking entropies $S_H$, 
$S_B$ and $S_{BH}$ by \cite{EV}
\bea
\label{XVI}
&& S_H^2=-{(4\pi)^2 V \over \kappa^4}\left(\int d\theta d\phi B^{1t}
+ {V \over a^2}\right) ={(4\pi)^2 V^2 \dot A^2 \over \kappa^4} 
\ ,\nn
&& S_B=\pi a \int d\theta d\phi B^{{\rm ext}\,1t}
=\pi a E
\ ,\quad S_{BH}={4\pi V \over \kappa^2 a}\ ,
\eea
with space volume $V$ and the total energy $E$ defined by
\be
\label{XVII}
V=\int d\theta d\phi\sqrt{g} = 2\pi a^2\ ,\quad 
E=\int d\theta d\phi\sqrt{g} \rho \ ,
\ee
Then from FRW equations it follows  the  relation\cite{EV} 
reminding about two-dimensional CFT entropy \cite{Cardy} 
(so-called Cardy-Verlinde formula)
\be
\label{XVIII}
S_H^2 + S_{BH}^2 = 2 S_{BH}S_B\ .
\ee
The above equation (\ref{XVIII})  follows from the first 
equation in (\ref{XV}).

This equation can be rewritten  as
\be
\label{XVIIIb}
S_H^2 = S_B^2 - \left(S_{BH} - S_B\right)^2\ .
\ee
The well-known square root like in two-dimensional CFT entropy
appears when one evaluates Hubble entropy from above relation.
We should note that 
\be
\label{XIX}
\Phi=\int d\theta d\phi B^{1t}
\ee
is total magnetic flux. Then the external flux 
\be
\label{XX}
\Phi^{\rm ext}=\int d\theta d\phi B^{{\rm ext}\,1t}
\ee
corresponds to the total energy
\be
\label{XXI}
\Phi^{\rm ext}=- E\ .
\ee
Since Eq.({XVIIIb}) tells $S_B^2 - \left(S_{BH} 
 - S_B\right)^2\geq 0$, Eqs.(\ref{XVI}) and (\ref{XXI}) 
give somel bound between the external magnetic 
flux and the spacial volume $V$:
\be
\label{XXIb}
\left(\Phi^{\rm ext}\right)^2 \geq \left({4V \over \kappa^2 
a^2} + \Phi^{\rm ext}\right)^2\ .
\ee
We should note, however, the spacial volume $V$ is given from
the gravity side. 

In order to proceed further, we consider the radiation as a matter. 
Then the energy density $\rho$ and the pressure $p$ have the 
following forms:
\be
\label{XXII}
\rho=2p=\rho_0\e^{-3A}\ .
\ee
The FRW equations (\ref{XV}) are
\bea
\label{XXIII}
&& \dot A^2 + {\e^{-2A} \over l^2}={\kappa^2 \over 2}
\rho_0\e^{-3A} \nn
&& \ddot A + \dot A^2={\kappa^2 \over 4}\rho_0\e^{-3A} \ .
\eea
The solution of (\ref{XXIV}) is given by using an intermediate 
variable $\eta$ as follows,
\be
\label{XXIV}
a=l\e^A=a_0\sin^2\eta\ ,\quad 
t=a_0\left(\eta - {1 \over 2}\sin 2\eta\right)\ .
\ee
The maximum  raidus of the universe $a_0$ is given by
\be
\label{XXIVb}
a_0={\kappa^2 l^3 \over 2}\rho_0\ .
\ee
The universe starts with the big bang  
at $\eta=0$ ($t=0$) and the universe ends with the big crunch 
at $\eta=\pi$ ($t=\pi a_0$). Then the universe has the period of 
$\pi a_0$. As this is
Euclidean spacetime,  one can regard  period as the inverse of the 
temperature $T$ 
\be
\label{XXV}
T={1 \over \pi a_0}\ .
\ee
Since the energy $E$ (\ref{XVII}) is given by 
\be
\label{XXVI}
E=4\pi a^2 \rho_0 \e^{-3A}=4\pi l^3 \rho_0 a^{-1}\ ,
\ee
we find that the Bekenstein entropy $S_B$ is constant:
\be
\label{XXVII}
S_B=4\pi^2 l^3 \rho_0 = {8\pi^2 \over \kappa^2}a_0
= {8\pi \kappa^2}T\ .
\ee
Here  Eqs.(\ref{XXIVb}) and (\ref{XXV}) are used. 
When the radius of the universe is maximal ($a=a_0$),  
the Hubble entropy $S_H$ and the Bekenstein-Hawking 
entropy $S_{BH}$ are
\be
\label{XXXVIII}
S_H=0\ ,\quad S_{BH}={16\pi \kappa^2}T\ .
\ee
In this case CV formula becomes the identity.

The interesting question is: what happens in terms of  
Yang-Mills theory? We now evaluate the Yang-Mills action 
by substituting the classical solution. 
 Being in the Euclidean spacetime,  
 the action can be regarded as the free energy $F$ 
divided by the temperature $T$:
\be
\label{FT}
S_{\rm YM}={F \over T}\ .
\ee
 Using the obtained free energy, the Yang-Mills theory 
entropy may be evaluated. 

In  curved spacetime, the action 
(\ref{Y1}) has the following form:
\be
\label{Y1b}
S_{\rm YM}={1 \over 4g_{\rm YM}^2}\int d^3x \sqrt{g}
\sum_{a=1}^3\left(
F^a_{ij}F^{a\,ij} + 4g_{\rm YM}^2A^a_i J^{ai}\right) \ .
\ee
Then (\ref{Y2}) looks as:
\be
\label{Y2b}
S_{\rm YM}={1 \over 2g_{\rm YM}^2}\int d^3x \sum_{a=1}^3\left(
{1 \over \sqrt{g}}B^a_iB^{a\,i} 
+ 2g_{\rm YM}^2\sqrt{g} A^a_i J^{ai}\right) \ .
\ee
Even in  curved spacetime, the Bianchi identity 
(\ref{Y4}) is always satisfied since the Bianchi identity is 
a topological equation. On the other hand, 
the equation of motion  (\ref{Y3}) is modified as 
\be
\label{Y3cv}
0={1 \over \sqrt{g}}\epsilon^{kij}D_k\left(
{ B^a_j \over \sqrt{g}}\right) - g_{\rm YM}^2 J^{ai}\ .
\ee
If we define, however, $J^{ai}$, instead of (\ref{VIIIb}), 
\be
\label{Y3cv2}
{1 \over \sqrt{g}}\epsilon^{kij}D_k\left(
{ B^{{\rm ext}\,a}_j \over \sqrt{g}}\right) = g_{\rm YM}^2 J^{ai}\ ,
\ee
Eq.(\ref{Y3cv}) is always satisfied.

By using the equations of motion (\ref{Y3cv}) and (\ref{VIII}), 
the action (\ref{Y2b}) can be rewritten in the following form:
\be
\label{XXXIX}
S_{\rm YM}=-{1 \over 2g_{\rm YM}^2} \int d^3x {1 \over \sqrt{g}}
\sum_{a=1}^3 B^a_iB^{a\,i} 
=-{1 \over 2g_{\rm YM}^2}\int d^3x {1 \over \sqrt{g}}
\sum_{a=1}^3 B^{{\rm ext}\, a}_iB^{{\rm ext}\, a\,i} \ .
\ee
Then with the help of (\ref{IX}) and 
(\ref{XIV}) one gets
\be
\label{XXXX}
S_{\rm YM}=-{\kappa^4 \over 8g_{\rm YM}^2} \int d^t \e^{2A}
\left(\rho^2 + 2p^2\right)\ .
\ee
By further substituting the solution  (\ref{XXIV}) and 
using (\ref{XXII}), (\ref{XXIVb}) and (\ref{XXV}), we obtain 
the following expression for the action
\be
\label{XXXXI}
S_{\rm YM}= - {6\pi^2 \over g_{\rm YM}^2} 
T\int_0^\pi {d\eta \over \sin^2\eta}\ .
\ee
The integration in (\ref{XXXXI}) diverges. The divergence may be 
absorbed into the renormalization of the Yang-Mills coupling 
$g$:
\be
\label{XXXXII}
{1 \over g_R^2}\equiv {1 \over g_{\rm YM}^2} 
\int_0^\pi {d\eta \over \sin^2\eta}\ .
\ee
 Using the renormailized coupling constant $g_R$, the action 
has the following form:
\be
\label{XXXXIV}
S_{\rm YM}= - {6\pi^2 \over g_R^2}T\ .
\ee
 From the identification (\ref{FT}), the free energy $F$ is 
given by
\be
\label{XXXXV}
F=- {6\pi^2 \over g_R^2}T^2\ .
\ee
Then the entropy is 
\be
\label{XXXXVI}
{\cal S}=-{dF \over dT}={12 \pi^2 \over g_R^2}T\ .
\ee
If we identify 
\be
\label{XXXXVII}
{12 \pi^2 \over g_R^2}={8\pi \kappa^2}\ ,
\ee
the entropy (\ref{XXXXVI}) is equal to the 
Bekenstein entropy $S_B$ (\ref{XXVII}). 
As we  consider curved spacetime, the spacial 
volume $V$ is given by (\ref{XVII}). Therefore the 
identification (\ref{XXXXVII}) gives a bound for the 
entropy of the Yang-Mills theory with the external source 
$J^{ai}$ (\ref{Y3cv2}): 
\be 
\label{XXXXVIII}
S_H^2 + S_{BH}^2 = 2 S_{BH}{\cal S}\ .
\ee
As we are now considering the curved spacetime, $S_{BH}$ and 
$S_B$  (\ref{XXXXVIII}) is given, as in (\ref{XVI}), by 
the structures of spacetime where the Yang-Mills field lives. 
Using (\ref{XVI}), (\ref{XIX}) and the identification 
${\cal S}=S_B$,  the entropy in the Yang-Mills theory 
maybe expressed in terms of the magnetic flux. We can also rewrite 
Eq.(\ref{XXXXVIII}) in the form of (\ref{XVIIIb}): 
\be
\label{XVIIIbb}
S_H^2 = {\cal S}^2 - \left(S_{BH} - {\cal S}\right)^2\ ,
\ee
which gives the bound for the entropy ${\cal S}$:
\be
\label{XXXXIX}
{\cal S}^2 \geq S_H^2\ ,\quad {\cal S}^2 \geq 
\left(S_{BH} - {\cal S}\right)^2 \ .
\ee
If necessary the above YM entropy bound found for 
the particular thermal solution mapped to three-dimensional 
FRW cosmology maybe completely expressed in terms 
of gauge fields and sources. Moreover, it is clear 
that similar procedure maybe applied to study 
new solutions and entropy bounds for other cases,
including YM theory in four and higher dimensions.

There are  
 recent gravitational theory results which have
the holographic origin: like FRW cosmology 
presentation in the form reminding about two-dimensional 
CFT entropy (see \cite{frw} for a recent review) 
or cosmological entropy bounds. The main lesson of
our investigation is that similar results  are in some sense universal.
They should be expected 
in other theories, like in the case of YM theory under consideration.

\end{document}